\documentclass[aps,pra,onecolumn,longbibliography,footinbib,twocolumn]{revtex4-1}  %

\usepackage{graphicx} 
\usepackage{float} 
\usepackage{dcolumn} 
\usepackage{bm,color}
\usepackage{amsmath,amssymb,dsfont,amstext,amsfonts}
\usepackage{extarrows}
\usepackage{xcolor}
\usepackage{wasysym}
\usepackage{mathtools}
\usepackage{bbold}
\usepackage[export]{adjustbox}
\usepackage{mathdots}
\usepackage{siunitx}
\usepackage[colorlinks=true,linkcolor=blue,urlcolor=blue,citecolor=blue]{hyperref}
\usepackage{mathrsfs}

\usepackage[normalem]{ulem} 
\usepackage{color}

\usepackage{multirow}
\usepackage[T1]{fontenc}
\usepackage{lmodern}

\usepackage{xcolor,colortbl}
\usepackage[english]{babel}
\usepackage{amsmath}
\usepackage{amssymb, comment}
\usepackage[]{graphicx}
\usepackage{times}
\usepackage{bm}
\usepackage{braket}
\usepackage[version=3]{mhchem}

\usepackage{bbm}
\usepackage{tabularx}

\usepackage{xspace}

\begin{document}
\title{Ultrafast anisotropic exciton transport in phosphorene}
\author{Kai-Wei Chang}
\email{kwc40@cam.ac.uk}
\affiliation{Department of Materials Science and Metallurgy, University of Cambridge, 27 Charles Babbage Road, Cambridge CB3 0FS, United Kingdom}

\author{Joshua J. P. Thompson}
\affiliation{Department of Materials Science and Metallurgy, University of Cambridge, 27 Charles Babbage Road, Cambridge CB3 0FS, United Kingdom}

\author{Bartomeu Monserrat}
\affiliation{Department of Materials Science and Metallurgy, University of Cambridge, 27 Charles Babbage Road, Cambridge CB3 0FS, United Kingdom}

\begin{abstract}
Phosphorene is a two-dimensional (2D) material exhibiting strong in-plane structural anisotropy. In this work, we investigate the influence of structural anisotropy on the optics, dynamics, and transport of excitons in phosphorene by combining microscopic many-body theory with first principles calculations. Our framework offers a complete and material specific description of the excitonic properties of phosphorene, including exciton states and exciton-phonon interactions, which allow us to quantitatively evaluate the optical absorption spectra, exciton relaxation, and exciton transport, revealing direction-dependent characteristics. Interestingly, we identify the critical role of long-range exchange interactions, which significantly enhance the anisotropy of exciton diffusion, particularly at low temperatures. Our work provides fundamental insights into exciton dynamics in an intrinsically anisotropic 2D material, offering guiding principles for the design of next-generation optoelectronic devices.

\end{abstract}
\maketitle

\section{Introduction}
Atomically thin two-dimensional (2D) materials have garnered significant attention over the past two decades due to their unique properties arising from dimensional confinement\,\cite{geim2007rise,manzeli20172d,akinwande2019graphene,das2021transistors}. Compared to their bulk counterparts, 2D materials exhibit distinct electronic, optical, mechanical, and thermal characteristics, making them highly suitable for next-generation ultrathin devices such as high-speed transistors\,\cite{lembke2012breakdown,liu2021promises}, flexible displays\,\cite{katiyar20232d}, and sensitive photodetectors\,\cite{long2019progress}, while also holding promise for revolutionary energy technologies including high-performance batteries\,\cite{el2016graphene,he2024layered}, supercapacitors\,\cite{el2016graphene,tomy2021emergence}, and efficient catalysts\,\cite{shifa2019heterostructures,deng2016catalysis}.

Coulomb-bound electron-hole pairs, known as excitons, play a fundamental role in the optical properties of semiconducting 2D systems\,\cite{perea2022exciton}. The reduced dimensionality and reduced dielectric screening in 2D materials compared to 3D materials enhance electron-hole interactions\,\cite{chernikov2014exciton}, resulting in large exciton binding energies and strong oscillator strengths\,\cite{zhang2018determination,jung2022unusually,zheng2019excitons,dvorak2013origin}. These features allow excitons to remain stable at room temperature\,\cite{wang2018colloquium,mak2010atomically} and they can be tuned via external fields\,\cite{klein2016stark,ross2013electrical}, strain\,\cite{yang2021strain,he2013experimental,castellanos2013local}, or environmental factors\,\cite{raja2017coulomb,lin2014dielectric}. This rich landscape of excitonic phenomena in 2D materials is not only of scientific interest but also central to the development of optoelectronic devices, including light-emitting diodes (LEDs)~\cite{ross2014electrically}, photodetectors\,\cite{long2019progress}, solar cells\,\cite{das2019role}, optical modulators\,\cite{yu20172d,sun2016optical}, and quantum light sources\,\cite{chakraborty2019advances}.


Anisotropy in materials offers a powerful means to control optical properties and energy transport\,\cite{thompson2022anisotropic}, opening new pathways for device design and functionality. Phosphorene, a two-dimensional material with inherently large in-plane anisotropy, has emerged as a promising platform for exploring such effects. The anisotropic optical response of phosphorene has been widely observed\,\cite{brunetti2019optical, ghosh2017anisotropic, wang2015highly, lu2016light}, evident in linearly polarised optical absorption and the emergence of hyperbolic exciton-polaritons. However, despite these intriguing findings, the dynamics and transport behavior of excitons in phosphorene remain largely unexplored.

In this work, we propose that the intrinsic anisotropy of phosphorene leads to highly directional exciton energy transport. We use microscopic many-body particle theory\,\cite{brem2018exciton} parametrised with first principles calculations to investigate the exciton properties of phosphorene. 
We compute: (i) the exciton states and corresponding optical absorption spectra to elucidate their directional dependence; (ii) the exciton relaxation dynamics following photoexcitation; and (iii) the anisotropic exciton diffusion constants along distinct crystallographic directions, which are found to be dramatically affected by long-range exchange interactions. Our findings demonstrate that the inherent exciton anisotropy in phosphorene is robust against phonon scattering, opening the door for intrinsic directional control of energy in a device. This will pave the way for new applications including exciton highways for light harvesting applications, linearly polarised light-emitting diodes, and future excitonic circuitry.  


\section{Methods}
\label{sec: methods}

\subsection{First principles calculations}
We perform structural optimization, and calculate the electronic band structure and the macroscopic dielectric constant using density functional theory as implemented in the Vienna Ab-initio Simulation Package ({\sc vasp})\,\cite{kresse1996efficient} with the projector augmented wave method\,\cite{blochl1994projector,kresse1999ultrasoft}. We use a $16\times12\times1$ Monkhorst-Pack $\mathbf{k}$-grid to sample the Brillouin zone\,\cite{monkhorst1976special} and set the energy cut-off to $500$\,eV. To achieve an accurate treatment of the exchange-correlation energy, we adopt the hybrid HSE06 functional for structural optimization and electronic band structure calculations\,\cite{krukau2006influence}. We maintain a vacuum spacing greater than $15$\,\AA\@ to prevent interlayer interactions, and optimise the structure to achieve residual forces below $0.01$\,eV\,\AA$^{-1}$. 

Using {\sc wannier90}\,\cite{mostofi2014updated}, we construct a tight binding model parametrised by the HSE06 calculations. We use this tight binding model to finely sample the electronic bands around select $\mathbf{k}$-points, we then fit the resulting two-dimensional band energy surfaces to an analytical functional form using linear regression, which we then use to construct the associated Hessian matrix, and finally evaluate the electron and hole effective masses.

For the macroscopic dielectric constant, we adopt the Perdew-Burke-Ernzerhof (PBE) generalised gradient approximation (GGA) to the exchange-correlation functional\,\cite{PBE-exchange-correlation}. We use the finite difference method implemented in {\sc vasp} to compute the dielectric tensor. The effective monolayer thickness and correction of the 2D dielectric constant follows the methodology proposed in previous works\,\cite{laturia2018dielectric}. 

We use {\sc Quantum ESPRESSO}\,\cite{giannozzi2017advanced} to calculate the phonon dispersion and electron-phonon matrix elements of phosphorene. We adopt norm-conserving pseudopotentials in conjunction with the PBE exchange-correlation functional\,\cite{PBE-exchange-correlation}. For phonons, we construct the dynamical matrix from the calculated force constants to get phonon energies at arbitrary $\mathbf{q}$-points. We also use {\sc Quantum ESPRESSO} to calculate the interband dipole moment, defined as the probability of electron transitions between different energy bands, for light absorption calculations.

For electron-phonon interactions, the matrix elements have the following form:
\begin{equation}
\label{el-ph}
g^{\mathbf{k},j,\alpha}_{\mathbf{q}} = D^{\mathbf{k},j,\alpha}_{\mathbf{q}}\sqrt{\frac{\hbar^2}{2\rho\hbar\omega^{j}_{\mathbf{q}}}},
\end{equation}
where $\omega^{j}_{\mathbf{q}}$ is the phonon frequency associated with wave vector $\mathbf{q}$ and branch $j$, $\alpha$ is the electronic band index, and $\rho$ is the mass density of the material. In phosphorene, the optical phonon modes are relatively flat and excitons are highly localised in momentum space. As a result, the exciton–phonon interaction associated with optical modes can be effectively described by transitions involving a narrow range of phonon energies and momenta, and we employ a constant deformation potential with respect to the phonon wave vector $\mathbf{q}$. For acoustic modes, we find that the deformation potential scales linearly with phonon wave vector for small $\mathbf{q}$, in agreement with previous works\,\cite{selig2016excitonic}, that is, $D^{k,j,\alpha}_{\mathbf{q}}=D^{k,j,\alpha}_{\mathrm{AC}}\left|\mathbf{q}\right|$. Overall, we interpolate the electron-phonon coupling coefficients using $g\propto \sqrt{q}$ for acoustic linear modes and keep the value constant for optical modes. The acoustic quadratic mode characteristic of 2D materials corresponds to out-of-plane atomic vibrations which have vanishing electron-phonon coupling in centrosymmetric materials. Due to the computational cost, we account for the directional dependence by evaluating the first principles values along several high-symmetry directions, and then use linear interpolation for general directions.

\subsection{Wannier equation}
To model the exciton behavior, we use the Wannier equation to describe the electron-hole interaction\,\cite{kira2006many,haug2009quantum}:
\begin{align}\label{wannier_k}
\frac{(\hbar \mathbf{k})^2}{2m_\mathrm{r}} \phi^\mu(\mathbf{k}) - \sum_{\mathbf{q}} V_{\mathbf{q}} \phi^\mu(\mathbf{k} + \mathbf{q}) = E_\mu^{\mathrm{b}} \phi^\mu(\mathbf{k}),
\end{align}
where $m_\mathrm{r}$ is the reduced mass arising from the electron $m_\mathrm{e}$ and hole $m_\mathrm{h}$ effective masses, and $V_{\mathbf{q}}$ is the screened Coulomb electron-hole interaction with momentum difference $\mathbf{q}$ between electron and hole. The excitonic binding enegy $E^{\mathrm{b}}_{\mu}$ and wavefunction $\phi^\mu(\mathbf{k})$ for state $\mu$ can be found by diagonalising the above equation, which gives a hydrogen-like series of excitonic energy levels. The parameter $\mathbf{k}$ of the exciton wavefunction is the relative electron-hole momentum, $\mathbf{k}=\alpha\mathbf{k}_\mathrm{h} + \beta\mathbf{k}_\mathrm{e}$, where $\alpha=m_\mathrm{e}/(m_\mathrm{e}+m_\mathrm{h}))$ and $\beta=m_\mathrm{h}/(m_\mathrm{e}+m_\mathrm{h})$. 

The Coulomb interaction in the Wannier equation can be modelled using the two-dimensional Keldysh potential\,\cite{rytova2018screened,keldysh2024coulomb}:
\begin{align}
V_{|\mathbf{q}|} = \frac{e^2_0}{2\varepsilon_0A|\mathbf{q}|(\varepsilon_{\mathrm{background}} + \varepsilon^\parallel_{\mathrm{material}}d|\mathbf{q}|/2)},
\end{align}
where $\varepsilon_0$ is the vacuum dielectric constant, $\varepsilon_{\mathrm{background}}$ is the background screening, $d$ is the thickness of the material, and $A$ is the surface area per unit cell. Throughout this study, we consider phosphorene on a SiO$_2$ substrate with a background $\varepsilon_{\mathrm{background}}$ of 3.9\,\cite{mcpherson2003trends}, to simulate realistic experimental conditions. 

We treat $\varepsilon^\parallel_{\mathrm{material}}$ as the average of the dielectric constant in the two in-plane directions, and we compare the excitonic calculation results with those using the interpolation method for the Keldysh potential\,\cite{galiautdinov2019anisotropic}. The resulting excitonic wavefunction distributions are nearly identical.

The exciton-phonon scattering matrix elements\,\cite{brem2018exciton}:
\begin{align}\label{eq:ex-ph}
W^{\eta \eta'}_{\mathbf{Q} \mathbf{Q}'} = \frac{2\pi}{\hbar} \sum_{\pm j} \left| G^{\eta \eta'j}_{\mathbf{Q}' - \mathbf{Q}} \right|^2 A^{j, \pm}_{\mathbf{Q}' - \mathbf{Q}} \, \delta\left( \Delta E^{\eta \eta'}_{\mathbf{Q'} - \mathbf{Q}} \pm \hbar \omega^{j}_{\mathbf{Q'} - \mathbf{Q}} \right),
\end{align}
describe scattering between exciton state $\eta$ with centre-of-mass momentum $\mathbf{Q}$ to the exciton state $\eta'$ with centre-of-mass momentum $\mathbf{Q'}$. In this expression, $G^{\eta \eta'j}_{\mathbf{Q} - \mathbf{Q}'}$ denotes the exciton-phonon scattering coefficient, which has the form $\sum_\mathbf{k}\phi_{\mathbf{k}}^{\eta~*}(\phi_{\mathbf{k}+\beta\mathbf{q}}^{\eta'~}g^{\mathbf{k},j,c}_{\mathbf{q}} - \phi_{\mathbf{k}-\alpha\mathbf{q}}^{\eta'}g^{\mathbf{k},j,v}_{\mathbf{q}})$, where $c$ and $v$ denotes the conduction and valence band, respectively; $A^{\nu, \pm}_{\mathbf{Q} - \mathbf{Q}'} = (\delta_{\pm1,1}+n^{\nu}_{\mathbf{Q} - \mathbf{Q}'})$ describes emission ($+$) and absorption ($-$) of a phonon; and the $\delta$ function imposes the exciton-phonon scattering channel, which in numerical calculations is replaced by a Lorentzian. The use of the Lorentzian can be justified by noting that truncating the semiconductor Bloch equation at higher order (including two-phonon processes) softens the usual delta function, giving rise to a self-consistent exciton-phonon scattering rate\,\cite{meneghini2024excitonic}. This expression for the exciton-phonon scattering rate can be derived using the semiconductor Bloch equations\,\cite{kira2006many}.

The dephasing describes the total scattering rate of an exciton in a specific state\,\cite{brem2018exciton}:
\begin{align}\label{eq:w_ex-ph}
\Gamma^{\eta }_{\mathbf{Q}} = \frac{2\pi}{\hbar} \sum_{\mathbf{Q'},\eta',\pm, j} \left| G^{\eta \eta'j}_{\mathbf{Q}' - \mathbf{Q}} \right|^2 A^{j, \pm}_{\mathbf{Q}' - \mathbf{Q}} \, \delta\left( \Delta E^{\eta \eta'}_{\mathbf{Q'} - \mathbf{Q}} \pm \hbar \omega^{j}_{\mathbf{Q'} - \mathbf{Q}} \right),
\end{align}
arising from the summation of the exciton-phonon scatterings with all other excitons of momenta $\mathbf{Q}'$ in state $\eta'$.

\section{Results and Discussion}
\label{sec: results}
\subsection{Crystal structure and physical parameters}

\begin{figure*}[!t]
\centering
\includegraphics[width=0.9\linewidth]{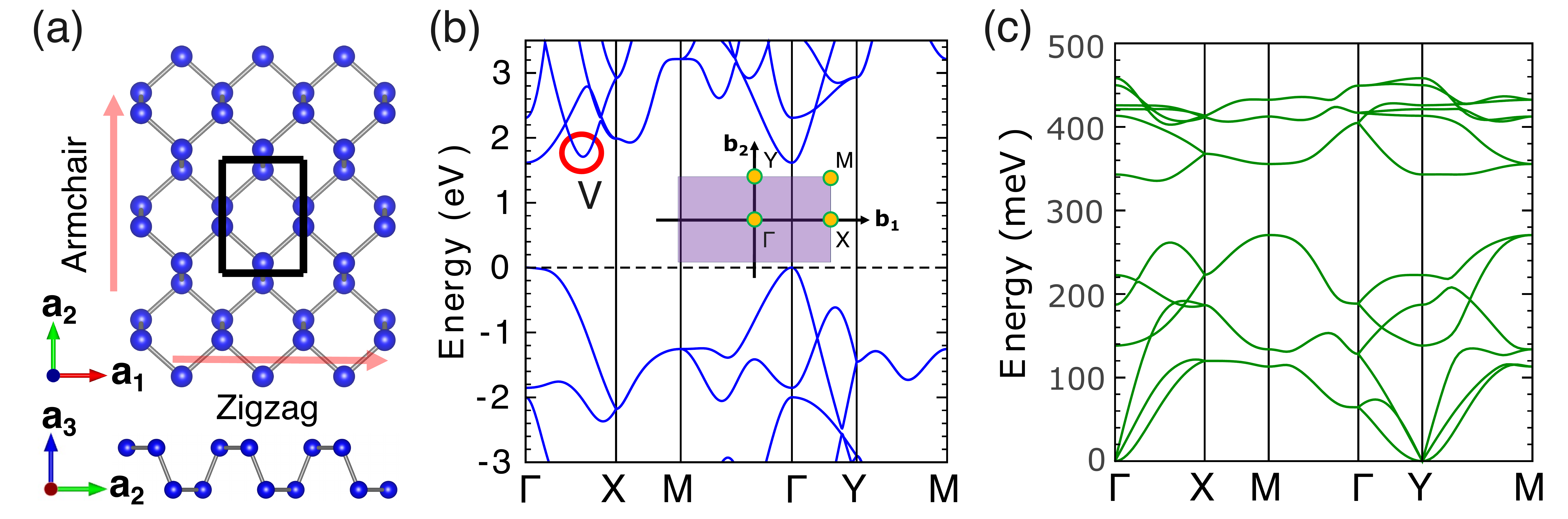} 
\caption{(a) Top view and side view of phosphorene. The directions along $\mathbf{a}_1$ and $\mathbf{a}_2$ are referred to as zigzag and armchair directions, respectively. (b) HSE06 band structure of phosphorene. The valley along $\Gamma$-X is labeled V throughout this work. The Brillouin zone is shown in the inset.  (c) Phonon spectrum of phosphorene.}
\label{fig:struct_bnd}
\end{figure*}

Figure \ref{fig:struct_bnd}(a) shows the crystal structure of phosphorene, a single layer of black phosphorous. Unlike the completely planar structure of graphene, phosphorene has a buckling structure. The primitive cell of phosphorene adopts an anisotropic rectangular shape and contains four phosphorous atoms. The calculated lattice constants of phosphorene are $\mathbf{a}_1=3.277$\,\AA\@ and $\mathbf{a}_2=4.576$\,\AA, and the calculated effective monolayer thickness is $5.3$\,\AA. By comparison, the experimental lattice constants of bulk black phosphorous are  $\mathbf{a}_1=3.31$\,\AA, $\mathbf{a}_2=4.38$\,\AA, and $\mathbf{a}_3=10.5$\,\AA, and the effective thickness of a single layer is $5.25$\,\AA~\cite{brown1965refinement}. 

In this work, we follow the convention to name $\mathbf{a}_1$ and $\mathbf{a}_2$ as armchair (AC) and zigzag (ZZ) directions, respectively.

The calculated macroscopic dielectric constants are $\epsilon_{\parallel(\mathrm{ZZ})}=12.17$,   $\epsilon_{\parallel(\mathrm{AC})}=16.13$, and $\epsilon_{\perp}=6.73$. 

The calculated band structure of phosphorene is shown in Fig.\,\ref{fig:struct_bnd}(b). It exhibits strong anisotropy around the valence band maximum (VBM), 
with the effective masses along the $\Gamma$–X (ZZ) and $\Gamma$–Y (AC) directions being $3.457\,m_0$ and $0.849\,m_0$, respectively, where $m_0$ is the free electron mass. For the conduction band minimum (CBM), the anisotropy is less pronounced: at the $\Gamma$ valley the effective masses along the ZZ and AC directions are $0.963\,m_0$ and $1.158\,m_0$, respectively; and around the secondary CBM valley (denoted as V), the effective masses are $0.272\,m_0$ and $0.220\,m_0$ for the ZZ and AC directions, respectively.


Figure~\ref{fig:struct_bnd}(c) shows the phonon dispersion of phosphorene. The lowest energy acoustic mode is the flexural ZA mode\,\cite{jiang2015review}, exhibiting the quadratic dispersion characteristic of 2D materials. Tables~\ref{tab:elph_cbm} and~\ref{tab:elph_vbm} list the calculated electron-phonon coupling coefficients. The modes exhibiting the largest electron-phonon coupling coefficients are selected for exciton-phonon coupling calculations.

\begin{table*}[h!]
\centering
\begin{tabularx}{.99\textwidth}{X X X X X X X}
\rowcolor{lightgray}
\multicolumn{7}{c}{Electron-phonon Coupling Coefficients/Phonon Energy (CBM)} \\
&$\mathbf{\Gamma\rightarrow\Gamma}$
&
&$\mathbf{\textbf{V}\rightarrow\textbf{V}}$
&
&$\mathbf{\Gamma\rightarrow\textbf{V}}$ \\  \toprule
\textbf{Phonon Mode}& $g_c$ & $E_{\mathrm{ph}}$ &  $g_c$ & $E_{\mathrm{ph}}$ &$g_c$ & $E_{\mathrm{ph}}$ \\ \toprule
Acoustic 1
&$\sim 0$
&
&$\sim 0$
&
&$10.7$
&$10.5$
\\
Acoustic 2
&$[\sim 0,\,11.3,\,\sim 0]$
&$[5.1,\,15.7,\,\sim 0]$
&$[\sim 0,\,9.6,\,6.6]$
&$[5.1,\,15.7,\,\sim 0]$
&$\sim 0$
&
\\
Acoustic 3
&$[44.1,\,31.3,\,31.2]$
&$[10.4,\,8.8,\,5.1]$
&$[39.4,\,30.0,\,18.1]$
&$[10.4,\,8.8,\,5.1]$
&$\sim 0$
&
\\
Optical 3
&$\sim 0$
&
&$\sim 0$
&
&$92.8$
&$32.4$
\\
Optical 4
&$116.0$
&$42.5$
&$\sim 0$
&
&$\sim 0$
&
\\
Optical 8
&$105.0$
&$55.8$
&$131.0$
&$55.8$
&$\sim 0$
& 
\\ \toprule
\end{tabularx}
\caption{Electron-phonon coupling coefficients $g_c$ and the corresponding phonon energies $E_\mathrm{ph}$ for selected phonon modes coupling with CBM states. The electron-phonon coupling coefficients show anisotropy around $\mathbf{q}=0$, so we use the format $[g_x/\sqrt{q},\,g_{xy}/\sqrt{q},\,g_y/\sqrt{q}]$ for this case. For the optical and intervalley acoustic modes, the units are given by meV, and for intravalley acoustic modes, the units are given by meV\,nm$^{-1/2}$. The units of phonon energy are meV.
}
\label{tab:elph_cbm}
\end{table*}

\begin{table*}[h!]
\centering
\begin{tabularx}{.99\textwidth}{X X X}
\rowcolor{lightgray}
\multicolumn{3}{c}{Electron-phonon Coupling Coefficients/Phonon Energy (VBM)} \\
&$\mathbf{\Gamma\rightarrow\Gamma}$
\\  \toprule
\textbf{Phonon Mode}
&$g_v$
&$E_{\mathrm{ph}}$
\\ \toprule
Acoustic 3
&$[14.9,\,7.9,\,3.1]$
&$[10.4,\,8.8,\,5.1]$
\\
Optical 4
&$6.4$
&$42.5$
\\
Optical 8
&$73.2$
&$55.8$
\\ \toprule
\end{tabularx}
\caption{Electron-phonon coupling coefficients $g_v$ and the corresponding phonon energies $E_\mathrm{ph}$ for selected phonon modes coupling with VBM states. The format of data and the units are the same as in Tab.\,\ref{tab:elph_cbm}.
}
\label{tab:elph_vbm}
\end{table*}

\subsection{Exciton landscape and optical absorption}
\begin{figure}[!t]
\centering
\includegraphics[width=0.8\linewidth]{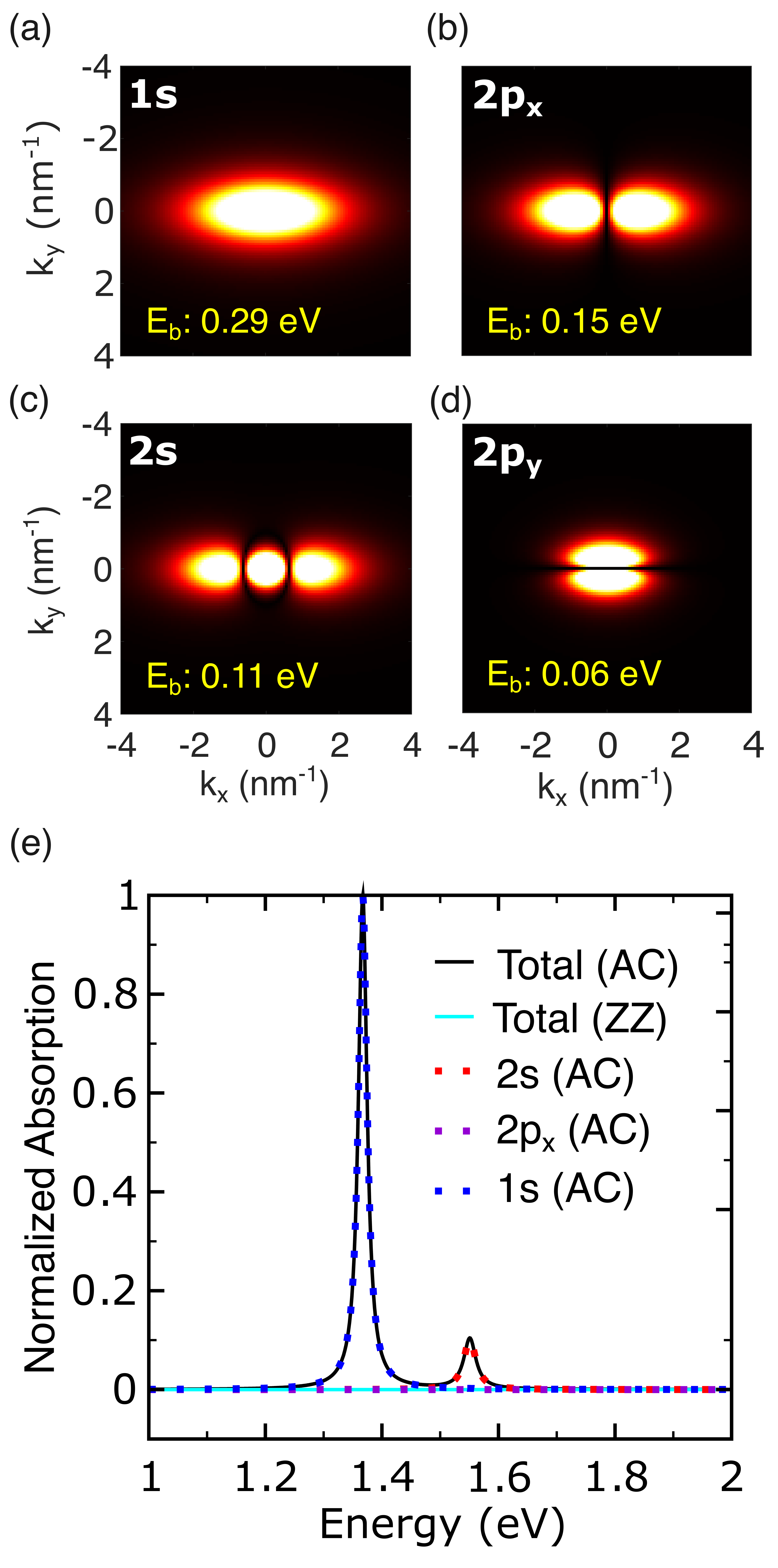} 
\caption{Spatial distributions of the (a) $1s$, (b) $2p_x$, (c) $2s$, and (d) $2p_y$ states of bright $\Gamma\Gamma$ excitons, with corresponding binding energies written in yellow. (e) Calculated light absorptions for the first three bright exciton states. The total absorptions along the AC and ZZ directions are depicted with solid lines, and contributions from individual states with dashed lines. Note that the absorption vanishes when the light polarization is along the ZZ direction.}
\label{fig:absorp_exciton}
\end{figure}

Figure~\ref{fig:absorp_exciton}(a-d) shows the calculated spatial distribution $|\phi_\mu^2|$ of the direct $\Gamma\Gamma$ excitonic wavefuntions obtained by solving the Wannier equation. The anisotropy of phosphorene leads to the 1s state exhibiting an elliptical shape in momentum space, which is consistent with the band structure shown in Fig.\,\ref{fig:struct_bnd}(b). Specifically, the electronic band structure has a broader band dispersion along $\Gamma$-X (the ZZ direction), leading to a more extended exciton wavefunction in this direction. Similarly, the spatial distribution of the $2p_{x}$ state is broader than that of the $2p_{y}$ state. Additionally, the anisotropy lifts the degeneracy of the phosphorene $2p_{x}$ and  $2p_{y}$ excitons, a scenario that is markedly distinct to the corresponding excitonic landscape in isotropic systems such as transition metal dichalchogenides. Notably, our calculated exciton binding energy of $E^{\mathrm{b}}_{1s}=0.29$\,eV for the $1s$ state is in a good agreement with the experimental value of $0.3$\,eV measured for phosphorene one SiO$_2$\,\cite{yang2015optical}.

The characteristic linear polarization of excitons in phosphorene is determined by the interband dipole matrix elements connecting the VBM and CBM. The numerical evaluation of these matrix elements, combined with the associated optical selection rules, show that the transition dipole is oriented along the AC direction, with a magnitude of $6.862$\,nm$^{-1}$. 

Light absorption can be described by the Elliot formula\,\cite{koch2006semiconductor}, which is obtained from the semiconductor luminescence equations\,\cite{brem2018exciton} and derived from the Heisenberg equation of motion\,\cite{kira2006many}:
\begin{align}\label{eq:ex-ph}
I^{\sigma}_{\text{abs}}(\omega) =
\frac{2}{\hbar} \sum_{\eta} \text{Im} \frac{|\tilde{M}^{\sigma}_{\eta}|^2}{E^\mathrm{b}_{\eta} - \hbar\omega - i(\gamma^{\sigma}_{\eta} + \Gamma^{0}_{\eta})},
\end{align}
where on the left-hand-side, $I$ denotes the absorption intensity, and $\sigma$ represents the polarization of the absorbed light. On the right-hand-side, in the numerator, $\tilde{M}^{\sigma}_{\eta}$ is the optical matrix element, given by the expression $\tilde{M}^{\sigma}_{\eta} = \mathbf{d}_{\eta} \cdot \mathbf{\hat{e}}_{\sigma} \sum_{\mathbf{k}} \phi^{\eta}_{\mathbf{k}}$, where $\mathbf{d_{\eta}}$ is the interband dipole of the valley for the given state $\eta$, and $\mathbf{\hat{e}}_{\sigma}$ is the unit vector of incident light polarization. In the denominator, the first term $E^\mathrm{b}_{\eta}$ is the exciton binding energy for the exciton state $\eta$; the second term $\hbar\omega$ corresponds to the energy of the incident light; and the final term accounts for two broadening mechanisms: the optical broadening $\gamma^{\sigma}_{\eta}$ arising from radiative decay, and the phonon-induced broadening $\Gamma^{0}_{\eta}$ which represents the dephasing of state $\eta$ at zero centre-of-mass momentum. 
Indirect $\Gamma$V excitons are optically dark due to large momentum transfer and hence do not couple to light. 

The calculated interband dipole indicates that the dipole is aligned along the AC direction, and light absorption will be greatly suppressed when the polarization of the incident light is along the ZZ direction which is perpendicular to the AC direction. Figure~\ref{fig:absorp_exciton}(e) shows the absorption spectrum of phosphorene, with a nearly vanishing absorption when light polarization is along the ZZ direction. There are two visible peaks when the light polarization is parallel to the interband diople (the AC direction), which stem from the $1s$ and $2s$ exciton states. In contrast, since the parity of $p_x$ is odd, the absorption of state $2p_x$ is forbidden by symmetry. The sum $\sum_{\mathbf{k}} \phi^{\eta}_{\mathbf{k}}$ in $\tilde{M}^{\sigma}_{\eta}$ is zero due to the odd parity. These anisotropic results are in good agreement with previous works~\cite{xiao2017excitons}.

\subsection{Phonon-mediated exciton dynamics}

Exciton dynamics, the time evolution of excitons scattering between states mediated by phonons, is a key factor in understanding exciton linewidths, relaxation, and transport.

\subsubsection{Exciton dispersion and linewidth}

\begin{figure}
\centering
\includegraphics[width=0.9\linewidth]{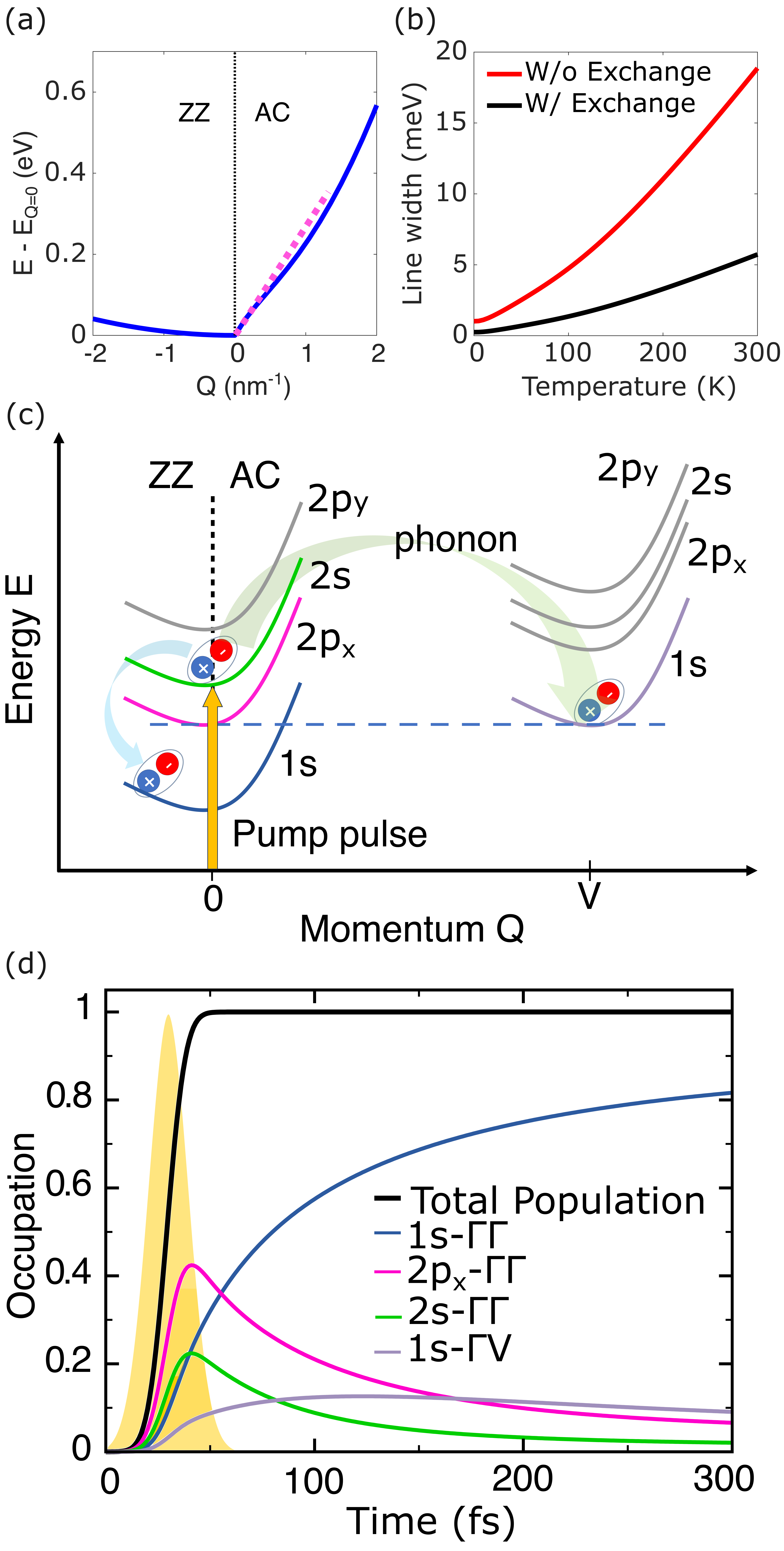}[!t]
\caption{ (a) Total exciton energy of the 1s state. The figure is divided into two parts: the left shows the ZZ direction, and the right shows the AC direction. The band distortion indicated by the pink dashed line shows the effect of the exchange energy. (b) Linewidth $\Gamma_{\mathbf{Q}=0}$ with and without exchange interaction. The red and black curves correspond to results without and with exchange interaction, respectively. (c) Schematic illustration of the exciton relaxation cascade. (d) Time evolution of the incoherent exciton population. The light pulse is shown with orange shadow area, and the total population is shown with solid line. Individual population of each states is demonstrated with different colours, as indicated in the legend.}
\label{fig:cascade_exchange}
\end{figure}

In studying exciton dynamics, an accurate description of the exciton energy is needed in order to reliably track the evolution of exciton states. The excitonic dispersion can be described as a parabolic dispersion, stemming from the electronic and hole valleys, and a $\mathbf{Q}$-dependent long-range exchange interaction $K_\mathbf{Q}$:
\begin{align}\label{eq:excitonEn}
E_{\mathbf{Q}}=\dfrac{\hbar^2\mathbf{Q}^2}{2M} -E_\mathrm{binding}+E_{\mathrm{gap}}+K_\mathbf{Q},
\end{align}
where $M = m_e+m_h$ is the total mass of the exciton. The long-range exchange interaction resembles a dipole-dipole interaction and has the form\,\cite{thompson2022valley}:
\begin{align}\label{eq:ex-ph}
K_{\mathbf{Q}} = \frac{V_{\mathbf{Q}}}{A} \left( \sum_{\mathbf{k}} \phi(\mathbf{k}) \mathbf{Q} \cdot \mathbf{d}_{\mathbf{k}+\mathbf{Q}} \right) \!\!\! \left( \sum_{\mathbf{k}'} \phi(\mathbf{k}') \mathbf{Q} \cdot \mathbf{d}_{\mathbf{k}'-\mathbf{Q}} \right),
\end{align}
where $A$ is the area per unit cell, and $\mathbf{d}_{\mathbf{k}}^{\mathrm{valley}}$ denotes the interband dipole matrix element within a given valley. We include the exchange interaction only for the $\Gamma\Gamma$ excitons, since the large momentum mismatch in $\Gamma$V excitons suppresses the long-range interaction, and we neglect the short-range exchange interaction as it is typically weak in two-dimensional materials\,\cite{thygesen2017calculating}. The Keldysh potential takes the form $V(\mathbf{Q})\approx c/(a\mathbf{Q}^2+b\mathbf{Q})$, implying that for small $\mathbf{Q}$, the linear term $b\mathbf{Q}$ dominates and $K_{\mathbf{Q}}$ scales linearly with $\mathbf{Q}$. For large $\mathbf{Q}$, $K_\mathbf{Q}$ tends to a constant, and the energy dispersion adopts to parabolic form driven by the shape of the electron and hole valleys. 

Figure~\ref{fig:cascade_exchange}(a) shows the energy dispersion of the $1s$ state of the $\Gamma\Gamma$ excitons. The intrinsic anisotropy of the effective masses leads to more dispersive excitonic energy bands along the AC direction compared to the ZZ direction. Moreover, since the dipole moment $\mathbf{d}$ is oriented along the AC direction, the scalar product $\mathbf{Q} \cdot \mathbf{d}$ vanishes in the ZZ direction, further enhancing the anisotropy in the energy dispersion. As discussed above, the pink dashed line of Fig.\,\ref{fig:cascade_exchange}(a) highlights how the original parabolic energy band is distorted by the exchange-induced linear term for small $\mathbf{Q}$, but exhibits the expected parabolic profile at larger $\mathbf{Q}$. The exchange contribution for small $\mathbf{Q}$ significantly affects exciton diffusion, as will be discussed later.

We show the linewidth $\Gamma_0^{1s}$ in Fig.\,\ref{fig:cascade_exchange}(b), which is significantly different when the long-range exchange interactions are included or excluded. Including the exchange interaction significantly reduces the linewidth because the linear dispersion at low $\boldsymbol{Q}$ makes scattering from $\boldsymbol{Q}=0$ less likely, owing to the reduced excitonic  density of states and more restrictive energy and momentum conservation.

\subsubsection{Exciton relaxation}

Figure~\ref{fig:cascade_exchange}(c) provides a schematic illustration of the exciton relaxation cascade. The final thermalised exciton population is independent of the initial exciton state that is excited. In our calculations, we choose to excite coherent excitons in $2s$ states by a resonant light pulse in order to illustrate how excitons relax from a higher to lower energy states. During the relaxation process, coherent excitons in $2s$ states are scattered by phonons to other incoherent exciton states\,\cite{brem2018exciton}. This polarization-to-population transfer is modeled by the following two microscopic differential equations:
\begin{align*}
\partial_t P^\eta = \left( \frac{i}{\hbar} E_\eta^{\mathrm{b}} - \gamma_\eta - \Gamma_\eta \right) P^\eta + i \frac{e_0}{\hbar m_0} \mathbf{M}_\eta \cdot \mathbf{A}_\sigma(t) \\
\partial_t N_{\mathbf{Q}}^\eta = \sum_{\eta', \mathbf{Q}'} \left( W_{\mathbf{Q}'\mathbf{Q}}^{\eta'\eta} N_{\mathbf{Q}'}^{\eta'} - W_{\mathbf{Q}\mathbf{Q}'}^{\eta\eta'} N_{\mathbf{Q}}^{\eta} \right) + \sum_{\eta'} W_{0\mathbf{Q}}^{\eta'\eta} |P^\eta|^2.
\end{align*}
The first equation describes the process of creating a coherent excitonic polarization, where $P^{\eta}\equiv \langle X_{\eta,\mathbf{Q}= 0}^\dagger \rangle$ is the excitonic polarization driven by a Lorentzian-like vector field $\mathbf{A}_\sigma(t)$ representing the laser pulse. We assume that the light polarization is parallel to the interband dipole. The second equation describes the exciton dynamics of population $N^{\eta}_\mathbf{Q}$ across exciton state $\eta$ at momentum $\mathbf{Q}$. The first term on the right-hand side determines the population transfer among exciton states, which will lead to the thermal equilibration among exciton states at long times. The final term accounts for the incoherent population generated from coherent excitons via exciton-phonon scattering. 

As shown in Fig.\,\ref{fig:cascade_exchange}(d), a $60$\,fs Lorentzian pulse excites the $2s$ state of the $\Gamma\Gamma$ exciton, which also increases the population of all other exciton states via phonon scattering. We do not consider the recombination of the electron-hole pair forming the exciton, so the total exciton population saturates after the pulse. At the beginning, the populations of $2p_x$, $2s$ and $1s$ states of $\Gamma\Gamma$ excitons grow. In this regime, the scattering from the $2s$-$\Gamma\Gamma$ state into the $2p_x$-$\Gamma\Gamma$ exciton is much faster than that into the $1s$-$\Gamma\Gamma$ exciton because the energy difference between the final $2p_x$-$\Gamma\Gamma$ state and the initial $2s$-$\Gamma\Gamma$ state is approximately $40$\,meV at $\mathbf{Q}=0$, which closely matches the phonon energy of the third optical phonon mode, causing stronger intravalley scattering matrix elements for this transition compared to the $2s$ to $1s$ transition. From about $40$\,fs, the subsequent dynamics between all states leads to the populations of the $2p_x$ and $2s$ states starting to fall, while that of the 1s state continues to rise, and its population surpasses those of the $2p_x$ and $2s$-$\Gamma\Gamma$ excitons around $80$ and $170$\,fs, respectively. As the system relaxes towards thermal equilibrium, the long-time population becomes largest for the lowest energy $1s$-$\Gamma\Gamma$ state, while the populations of the $2p_x$ $\Gamma\Gamma$-exciton and the $1s$ $\Gamma$V-exciton converge to similar values beyond $300$\,fs because the energies of these states are very similar. The $2s$ $\Gamma\Gamma$-exciton has the lowest population since its exciton energy is higher than that of other states. Finally, we omit the $2p_x$ and $2s$ states of the $\Gamma\mathrm{V}$ exciton in the figure due to their negligible contributions. We refer the reader to Fig.\,\ref{fig:absorp_exciton} and Tables~\ref{tab:elph_cbm} and~\ref{tab:elph_vbm} for the excitonic binding energies, electron-phonon coupling coefficients, and phonon mode energies of the system that drive the above relaxation processes.

\begin{figure}[!t]
\centering
\includegraphics[width=0.95\linewidth]{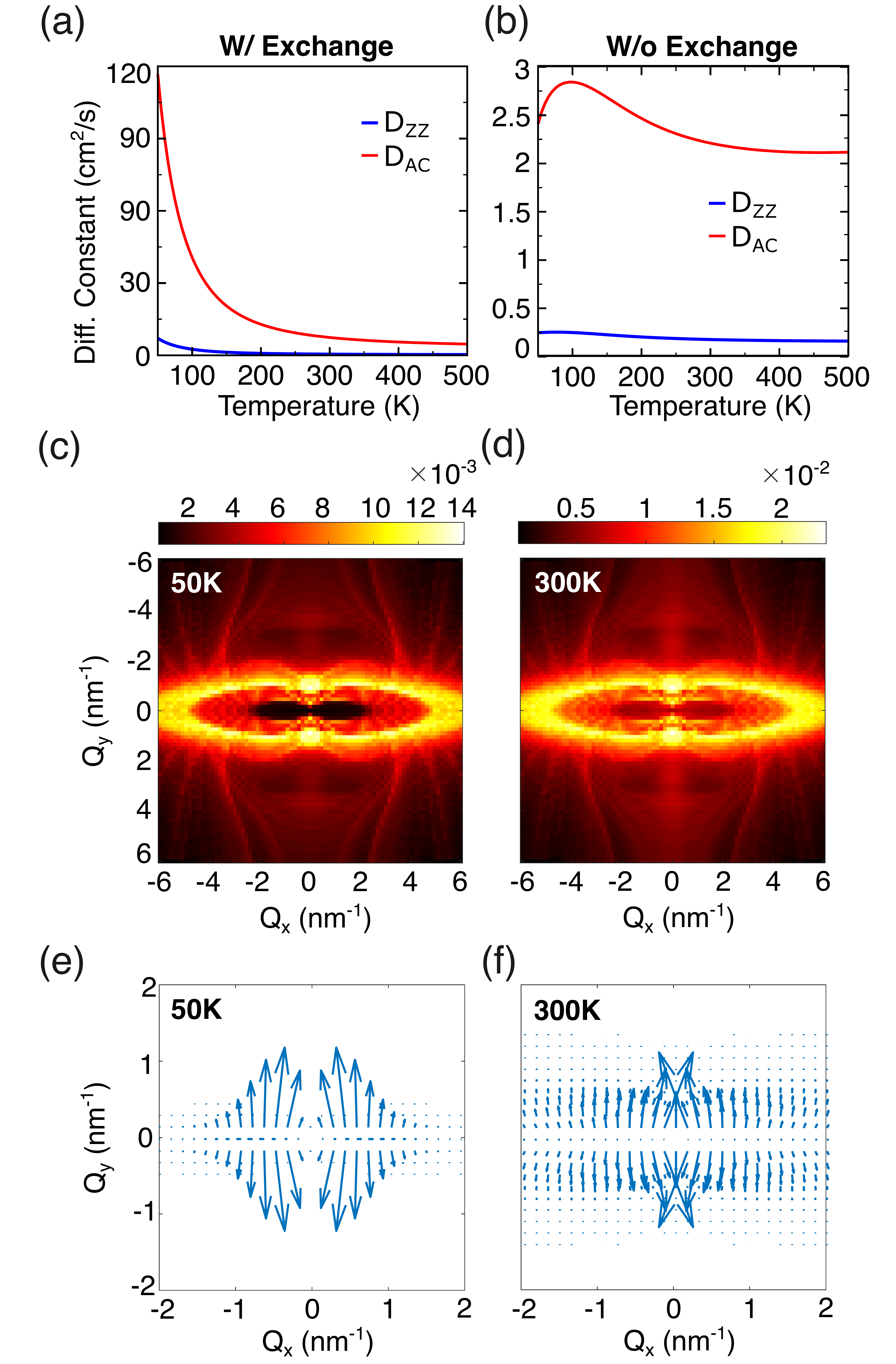} 
\caption{Calculated diffusion coefficients of phosphorene. (a) With exchange interaction (b) Without exchange interaction. The red and blue curves correspond to AC and ZZ directions, respectively. The 2D dephasing strength at (c) 50 K and (d) 300 K. Thermal expectation values of $|v_{g}|^2\cdot \hat{v_{g}}$ at (e) 50 K and (f) 300 K. Arrows indicate the direction of $\hat{v_{g}}$.}
\label{fig:diff}
\end{figure}

\subsubsection{Exciton transport}

Exciton transport can be described by tracking the spatiotemporal evolution for an initial exciton distribution. In an anisotropic system, this is well-described by Fick's second law of diffusion\,\cite{fick1855ueber, thompson2022anisotropic}, and is given by $\partial_t N(\mathbf{r}) = (D_{x} \partial_x^2 + D_{y} \partial_y^2) N(\mathbf{r})$. In the formula, $t$ denotes time, $N$ is the exciton population, and $D_{i}$ represents component $i$ of the diffusion constant:
\begin{align}\label{eq:diff}
D_{i} = \sum_{\mathbf{Q}\eta} \frac{(v_\mathbf{Q}^{\eta,i})^2}{2\Gamma_\mathbf{Q}^{\eta}} \frac{e^{-E_{\mathbf{Q},\eta}/k_B T}}{\mathcal{Z}},
\end{align}
where $v_\mathbf{Q}^{\eta,i}$ is the group velocity associated with the exciton in state $\eta$ and centre-of-mass momentum $\mathbf{Q}$ along the $i\mathrm{th}$ direction, $\Gamma_\mathbf{Q}^{\eta}$ is the dephasing, $E_{\mathbf{Q},\eta}$ is the exciton energy, and $\mathcal{Z}$ is the partition function. 

In Fig.\,\ref{fig:diff}(a,b), we compare the diffusion constants with and without the inclusion of the exchange interaction. In both cases, $D_\mathrm{AC}$($D_{x}$) is much larger than $D_\mathrm{ZZ}$($D_{y}$), which can be rationalized through the associated band dispersions. Specifically, the diffusion coefficient is proportional to the square of the group velocity, which in turn can be expressed as $v^{\eta}_{\mathbf{Q}}=\frac{1}{\hbar}\nabla_{\mathbf{Q}}E_{\mathbf{Q},\eta}$, where $E_{\mathbf{Q},\eta}$ has the form in Eq.\,(\ref{eq:excitonEn}). This implies that the diffusion constant is inversely proportional to the squared value of the effective mass, explaining why the diffusion coefficient $D_\mathrm{AC}$ is much larger than $D_\mathrm{ZZ}$. 

Another important feature of the diffusion coefficients is their temperature dependence. As shown in Fig.\,\ref{fig:diff}(a, b), the diffusion coefficients (with or without exchange) decrease in the high temperature regime. This is driven by an increase in dephasing with rising temperature, leading to a reduction in the diffusion constants at high temperature. Figures~\ref{fig:diff}(c) and (d) show the dephasing landscapes at temperatures of $50$\,K and $300$\,K, respectively, illustrating that the values at $300$\,K are significantly greater than those at $50$\,K. In the opposite low-temperature limit, the value of the diffusion constant in the AC direction soars. This is driven by a combination of the fact that the exciton population accumulates around $\mathbf{Q}\approx0$ at low temperature, and the steep gradient of the energy dispersion for small $\mathbf{Q}$ due to exchange along the AC direction.

Figures~\ref{fig:diff}(e) and (f) show the landscapes of the thermal expectation value of $|v_\mathbf{Q}^{\eta}|^2\cdot\hat{v}$ at $50$\,K and $300$\,K, respectively. Exciton velocity vectors in these two figures evidently align more with the AC direction, and the magnitude is bigger with small $\mathbf{Q}$, especially at low temperature. This anisotropic enhancement in mobility has direct implications for temperature-tunable excitonic device design.

Finally, we note that the calculated values of the diffusion coefficient for the AC and ZZ directions at $300$\,K are 7.40 and 0.48 cm$^2$/s, with their average value similar to the experimental value for a $2$\,nm amorphous phosphorene sample at $5.0$\,cm$^2$/s\,\cite{bellus2017amorphous}.

\section{Conclusions}
\label{sec: vi}

We have investigated the excitonic behavior of phosphorene using a fully microscopic framework, with all essential parameters calculated from first principles. Our results illustrate the large impact that the intrinsic structural anisotropy of phosphorene has on its excitonic properties. The anisotropy in optical absorption originates from the directionality of the interband dipole moment, which aligns along the armchair (AC) direction, in good agreement with experimental observations. For exciton transport, our calculations reveal that both exciton effective masses and long-range exchange interactions significantly contribute to highly different diffusion constants in the two in-plane directions, with exchange making a significant contribution in enhancing diffusion in the AC direction. Our findings unveil the intrinsic anisotropy in the excitonic behavior of phosphorene.

\begin{acknowledgments}
J.J.P.T. and B.M. acknowledge support from a EPSRC Programme Grant [EP/W017091/1]. K.-W.C. and B.M. acknowledge support from a UKRI Future Leaders Fellowship [MR/V023926/1]. B.M. also acknowledges support from the Gianna Angelopoulos Programme for Science, Technology, and Innovation. The computational resources were provided by the Cambridge Tier-2 system operated by the University of Cambridge Research Computing Service and funded by EPSRC [EP/P020259/1], and by the UK National Supercomputing Service, ARCHER. Access to ARCHER was obtained via the UKCP consortium and funded by EPSRC Grant No. EP/P022561/1.
\end{acknowledgments}

\end{document}